# Broadband Ultrahigh-Resolution chip-scale Scanning Soliton Dual- Comb Spectroscopy

Tong Lin[1], Avik Dutt[3], Chaitanya Joshi[2], XingchenJi[1,4], Christopher T. Phare[1,4], Yoshitomo Okawachi[2], Alexander L. Gaeta[2], Michal Lipson[1,*]

[1]Department of Electrical Engineering, Columbia University, New York, New York 10027, USA

[2]Department of Applied Physics and Applied Mathematics, Columbia University, New York, New York 10027, USA

[3]Ginzton Laboratory and Department of Electrical Engineering, Stanford University, Stanford CA 94305, USA

[4]School of Electrical and Computer Engineering, Cornell University, Ithaca, New York 14853, USA

**Abstract**

We present a chip-scale scanning dual-comb spectroscopy (SDCS) approach for broadband ultrahigh-resolution spectral acquisition. SDCS uses $Si_3N_4$ microring resonators that generate two single soliton micro-combs spanning 37 THz (300 nm) on the same chip from a single 1550-nm laser, forming a high-mutual-coherence dual-comb. We realize continuous tuning of the dual-comb system over the entire optical span of 37.5 THz with high precision using integrated microheater-based wavelength trackers. This continuous wavelength tuning is enabled by simultaneous tuning of the laser frequency and the two single soliton micro-combs over a full free spectral range of the microrings. We measure the SDCS resolution to be 319±4.6 kHz. Using this SDCS system, we perform the molecular absorption spectroscopy of $H^{13}C^{14}N$ over its 2.3 THz (18 nm)-wide overtone band, and show that the massively parallel heterodyning offered by the dual-comb expands the effective spectroscopic tuning speed of the laser by one order of magnitude. Our chip-scale SDCS opens the door to broadband spectrometry and massively parallel sensing with ultrahigh spectral resolution.

Microresonator-based frequency combs, by virtue of their large bandwidth and narrow comb-tooth linewidth, could enable broadband high-resolution spectroscopy critical for applications, such as medical diagnostics, atmospheric chemistry, combustion, reaction kinetics, etc (1–3). Such frequency combs can contain several hundred thousand narrow spectral lines (4, 5), or "teeth", that are mutually phase coherent and equidistant in frequency with a separation $f_{rep}$. Dual-comb spectroscopy (DCS) leverages these advantages of bandwidth and phase coherence, enabling fast detection using a single photodetector. DCS uses a pair of frequency combs with a small difference in repetition rates $\Delta f_{rep}$. The first comb with $f_{rep}$ measures the optical response of samples while the second comb $f_{rep} + \Delta f_{rep}$ acts as a local oscillator. Dual-comb interferograms that encode the spectral information are obtained by spatial mixing two combs in one photodetector to realize multi-heterodyne detection in the down-converted RF domain without any moving parts.

Dual-comb spectroscopy based on microresonators could enable high-resolution broadband spectroscopy on-chip. However to date, the technique has been limited to a resolution of tens of GHz (6-13). This low resolution leads to sparse spectral sampling and arises due to the micro-combs' large $f_{rep}$ [~ free spectral range (FSR)] as a result of their intrinsically short cavity length (6–14). Attempts to improve the resolution using larger cavities with smaller FSRs inevitably increase soliton threshold power due to the larger propagation losses and smaller finesse (15, 16).

We introduce and experimentally demonstrate a new approach that addresses the sparse spectral sampling limitation of microresonator-based dual-comb spectroscopy for broadband, ultrahigh-resolution sensing. Our chip-scale scanning dual-comb spectroscopy (SDCS) technique is based on two high-Q silicon nitride ($Si_3N_4$) microrings and achieves high-resolution using integrated platinum microheaters for high-resolution scanning of the micro-combs over a full FSR. As shown in the schematic in Fig. 1(A), a continuous wave (CW) laser split equally by an on-chip multimode interference (MMI) coupler, pumps two microrings with slightly different radii. We use integrated microheaters to lock the two single-soliton micro-combs to a fixed pump cavity detuning by stabilizing soliton powers [Fig. 1(A) and Fig. S1(A)]. The dual-comb is generated after combining the two soliton micro-combs in a directional coupler. One of the dual-combs interacts with the sample and the other is used as a reference. For continuous DCS scanning over the entire comb bandwidth, the pump laser frequency is tuned between adjacent comb teeth and the two microring resonances follow simultaneously with high precision assisted by the microheater-based wavelength trackers [See (17) for details of the experimental setup]. By scanning the laser frequency across only one FSR ($\approx f_{rep}$), the entire optical bandwidth can be accessed in the RF domain.

All the photonic components on our $Si_3N_4$ platform (18) are lithographically defined on a single 1 mm×1mm chip [Fig. 1(B)]. $Si_3N_4$ has been shown to be an excellent platform for nonlinear photonics (19) due to its low linear loss (~0.8 dB/m at 1550 nm) over a wide transparency window (~ 3 eV), and ability to be dispersion engineered (20) due to its high refractive index (~2). The radii of the two microrings (~114 μm) used for DCS [R1 and R2 in Fig. 1(B)] are designed to be slightly different, generating a dual-comb with 197 GHz $f_{rep}$ and $\Delta f_{rep}$ (< 10 MHz) used in this work [highlighted in Fig. 1 (B)]. The loaded Q-factors of R1 and R2 are about $2.2\times10^6$, resulting in a full-width at half-maximum (FWHM) of 85 and 87 MHz respectively [Fig. S1(E)]. The cross section of the waveguide for tuning the microrings is 1550 nm wide and 750 nm tall with 2-μm-thick $SiO_2$ top cladding. This geometry ensures a low anomalous group-velocity dispersion (GVD) near the 1560 nm pump wavelength (9). Platinum microheaters are fabricated on each cladded microring. The microheaters regulate each cavity resonance by tuning the refractive index of the $Si_3N_4$ microring (dn/dT ~ $4\times10^{-5}$ RIU/K) via the $Si_3N_4$ thermo-optic effect. The rise and fall times of the microheaters are 20 μs and 35 μs respectively [Fig. S2(A)]. This fast response time is critical to implement a closed-loop feedback control and stabilization scheme, which enables single soliton micro-comb operations with high stability during the entire scanning period.

We use integrated microheaters for independent modelocking (21) and fast frequency tunability of the microrings. The two single solitons (R1 and R2) both span 37.5 THz (300 nm) and are generated using the same pump source (centered at 1557.2 nm), ensuring high mutual coherence. The optical spectra of these two single solitons are shown in Figs. 2(A) and (B). After mixing the two solitons on one photodetector, a sequence of beat notes is recorded in the RF domain corresponding to multiples of $\Delta f_{rep}$. Since the comb lines are symmetric in frequency relative to the pump, the RF beat notes on the red side of the pump overlap with those on the blue side. For one-to-one mapping of the optical domain onto the RF domain, we shift the pump frequency by 100 MHz using an acousto-optic modulator (AOM) to access the blue side of the dual-comb and use the long pass filter to access the red side of the dual-comb. The

RF comb lines are derived from the Fourier transform of the interferograms and are equidistantly spaced by $\Delta f_{rep}$ =3 MHz with a linewidth of 60 kHz [Fig. S3(A)] inherited from the free-running pump laser source noise. The small repetition-rate difference $\Delta f_{rep}$ enables mapping from the optical span of 37.5 THz into a RF span of 570 MHz. The central RF lines have a signal-to-noise ratio (SNR) around 28 dB limited by the amplified spontaneous emission noise of the erbium-doped fiber amplifier.

We tune each of the single soliton micro-combs, by concurrently tuning the pump laser wavelength and the two microring cavity resonances. To ensure stable soliton powers while scanning the pump laser frequency, we develop a microheater feedback system to maintain the same pump-cavity detunings for both combs [Fig. S1(A)]. This system also compensates the on-chip thermal crosstalk between R1 and R2 as we thermally tune R1/R2 resonances by one FSR. The wide tuning range of our microheater of at least one FSR enables us to reliably align each resonance to the same pump laser regardless of initial detunings caused by fabrication variations. As a measure of pump-cavity detuning, the system monitors the soliton power after filtering out the pump as they are linearly proportional (22, 23). A fraction of the output power (10%) of each single soliton is tapped and sent to two photodetectors respectively as inputs of two proportional-integral-derivative (PID) controllers. The PID outputs are fed back to the microheaters for stabilizing pump-cavity detunings. On sweeping the pump laser frequency, the cavity resonances of R1 and R2 follow synchronously, enabling rapid frequency scanning for SDCS. The fast microheater response time of 20 μs enables a servo bandwidth of several kHz, which is more than sufficient for tracking and compensating pump-cavity detuning fluctuations in laser frequency scanning.

We show the ability to cover the entire optical bandwidth from 1420 nm to 1720 nm with no missing gaps, while maintaining a nearly constant amplitude for each comb line. To characterize continuous tuning of the optical spectra of each single micro-comb, we acquire 400 optical spectra of the R2 soliton while sweeping the pump laser frequency with the microheater feedback enabled [Figs. S2(B) and (C)]. Figure 2 (E) plots the wavelength of each R2 soliton comb line generated using the pump laser, which is tuned from 1557.2 nm to 1558.8 nm and all comb lines are tuned simultaneously.

As a proof of concept demonstration of SDCS, we measure the absorption spectra of gas-phase $H^{13}C^{14}N$ and show ability to resolve its overtones in the near infrared. We pump at 1560.4 nm and filter out 12 comb lines using a fused fiber WDM filter (1550 ± 7.5 nm). A fraction of the dual-comb bandwidth (1540-1560 nm) is selected to measure the $2v_3$ vibrational band of $H^{13}C^{14}N$ in the near-infrared spectral region. The details of the experimental setup are in [Figs. S1(A) and (D)]. Figure 3(A) shows the $H^{13}C^{14}N$ absorption spectra measured with SDCS and the reference data provided by the manufacturer (24), in which the two curves are in good agreement. Figure 3(B) shows the zoomed-in measurement of the P20 lines of the $2v_3$ band. Note that the fine-tuning range of the pump laser driven by the piezo-actuator is about 31 GHz (~FSR/7), the scanning process currently requires seven interleaving spectra to cover the entire optical bandwidth. We tune the pump laser at a speed of 0.62 THz/s by sending a triangular voltage signal to the piezo-actuator of the laser. Each acquisition takes 50 ms for a 0.37 THz spectral span, which corresponds to an overall tuning speed of 7.4 THz/s, limited by the laser piezo-actuator and the WDM filter bandwidth.

To characterize the SDCS resolution, we measure the transmission of an inline fiber coupled Fabry-Perot interferometer (FSR~15 GHz, FWHM<50 MHz). We use the dual-comb employed in Fig.3 to generate a high-resolution spectra spanning 2.5 THz by scanning the pump from 1560.35 nm at the speed of 77.5 GHz/s [Fig. S1(A)]. In order to calibrate the frequency of the dual-comb, we measure the $H^{13}C^{14}N$ gas cell simultaneously. We sweep the pump laser frequency using the piezo-actuator and interleave 7 acquisitions in order to cover the entire bandwidth of the WDM filter. We obtain a minimum laser tunning step of 319±4.6 kHz, by averaging a 4.096-µs-long sample of 25 interferograms. In Fig. 4(A), we plot the measured transmission spectral features of the Fabry-Perot interferometer. One can see that all of the interference fringes are resolved. The right panel of Figure 4 shows a higher resolution view of one interference fringe near 1558 nm, which is close to the P20 branch line of $H^{13}C^{14}N$ [Fig. 3(C)]. One can see that the 33-MHz-wide FWHM of the fringe is fully resolved. The left panel of Fig. 4(B) shows the minimal spectral sampling interval of 319 ±4.6 kHz (2.55±0.036 fm).

In summary, we demonstrate a powerful, high resolution chip-scale approach for spectroscopy over large spectral bandwidths (37.5 THz) and high resolution (<400 kHz). Our technique can scan the entire comb spectra with a spectral sampling of <400 kHz approaching the comb-tooth linewidth. This scanning over only one FSR extends the mode-hop-free fine-tuning range of commercial CW lasers from tens of GHz to tens of THz (25, 26). In addition, the parallel measurement of all comb teeth increases the laser tuning speed by a factor of $N$ (the comb tooth number) without compromising resolution. The SNR can be further improved by coherent averaging (27) and real-time compensation (28). A frequency coverage of 0.37 THz was achieved by collecting about 2.4 million interferograms in a single shot over 400 ms at a sampling rate of 250 MS/s. The total scanning time for 2.5 THz is 2.8 s, which is limited by the laser's piezo-actuator and the passive filter bandwidth. The speed of SDCS can be greatly enhanced by either using laser-integrated Kerr combs (29) or by using non-tunable CW lasers with electro-optic modulators (30–32). SDCS can potentially generate dual-combs from microrings with a 10 GHz FSR (16), further increasing the acquisition speed. Our SDCS technique facilitates a portable system-on-a-chip for ultrahigh-spectral-resolution broadband parallel sensing. Going forward, we envision a fully integrated SDCS chip, where a laser comb (29) serves as the light source, germanium photodetectors are used (33) for microheater control, a field-programmable gate array (34) enables dual-comb initialization and automated scanning, and a spiral waveguide extends light-matter interaction with samples (35).

## Data availability

The data that support the findings of this study are available from the corresponding authors on reasonable request.

## Acknowledgments

We acknowledge fruitful discussions with A. Mohanty, S. Miller, and M. Yu from Columbia University for the DCS measurements. We thank M. C. Shin for designing the PCB circuit. This work was performed in part at the Cornell Nano Scale Facility, a member of the National Nanotechnology Coordinated Infrastructure, which is supported by the NSF (grant ECCS-1542081), in part at the City University of New York Advanced Science Research Center NanoFabrication Facility and in part at the Columbia Nano Initiative shared labs at Columbia University in the City of New York. We acknowledge support from the Air Force Office of Scientific Research (FA9550-15-1-0303), the Defense Advanced Research Projects Agency (HR0011-19-2-0014), Advanced Research Projects Agency-Energy (DE-AR0000843)

## Conflict of Interests

The authors declare no competing financial interest.

## Contributions

T.L. carried out the experiments and analyzed the data with assistance from A.D., C.J., C.T.P and Y.O. All authors discussed the results and conclusions. X.J. fabricated the devices. T.L. drafted the manuscript, which was revised by M.L., with inputs from all other authors. M.L. and A.L.G. supervised the project.

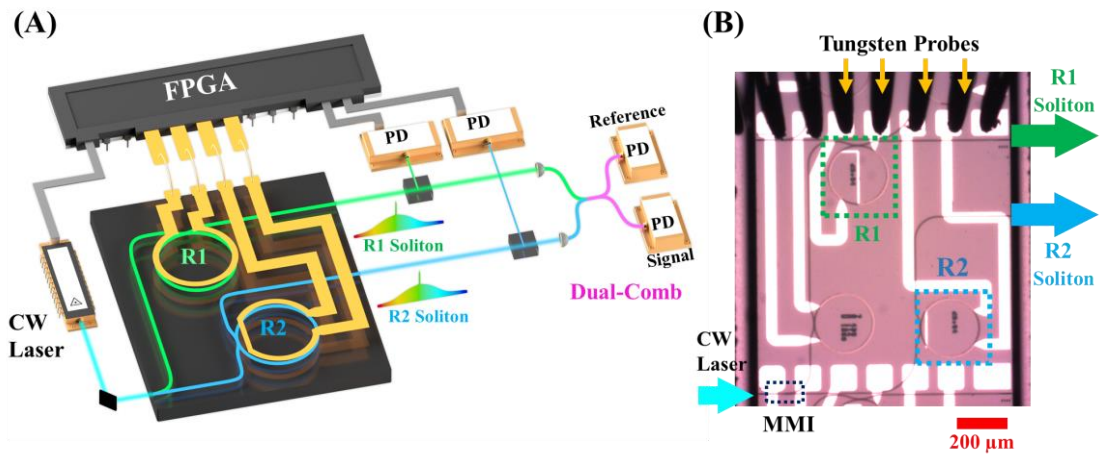

**Figure 1 | Chip-scale scanning dual-comb spectroscopy. (A)** Conceptual schematic of the system. A tunable CW laser is split into two via an MMI and used for pumping the two $Si_3N_4$ microring resonators. Through cascaded Four Wave Mixing and Kerr soliton formation, two single-soliton micro-combs are generated with two slightly different repetition rates: $f_{rep1}$ and $f_{rep2}$. Two individual single solitons are stabilized for a constant power by measuring a fraction of their power and feedbacking to microheaters atop. Microheaters are used to tune the comb spectra, with which cavity resonances are locked to the tunable laser frequency. The two soliton micro-combs are combined in a directional coupler to generate a dual-comb. One of the dual-comb is sent through a test sample. The other one provides a reference; **(B)** An optical microscope image of the fabricated device (top view) with integrated platinum microheaters contacted by tungsten probes. Light is coupled in and out of the chip through inverse tapers.

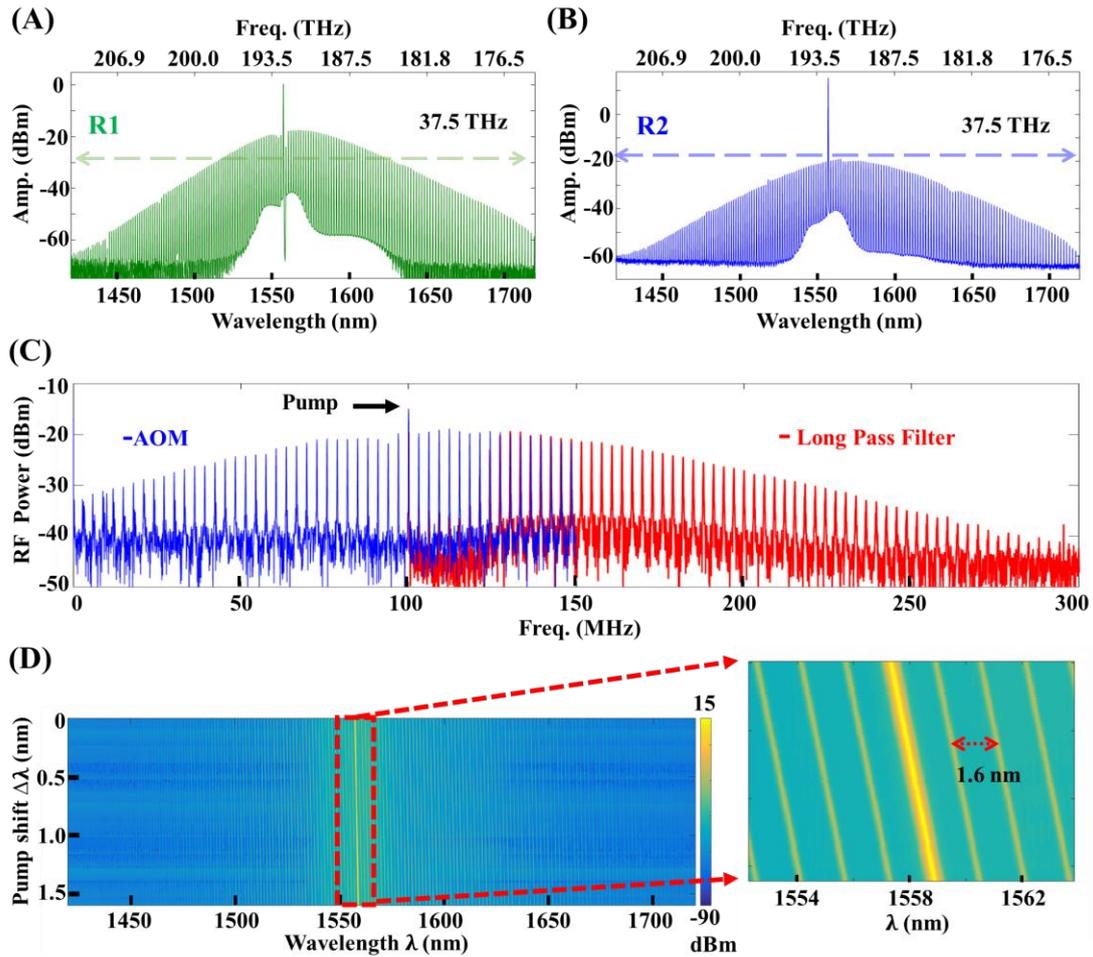

Figure 2 | Soliton micro-comb spectral characterization. (A and B) Optical spectra of microring R1 (with the pump filtered out by a fiber Bragg grating) and R2 single soliton micro-combs. (C) Resolved RF comb lines assisted by an AOM and a long pass filter (shift the spectrum to 100 MHz for matching the AOM carrier frequency). (D) The spectral tuning of the broadband R2 soliton micro-comb measured using an OSA with a resolution of 0.01 nm (the color bar denotes the optical power); the right panel shows the zoomed-in view of central seven comb lines, which are tuned over 1 FSR≈1.6 nm. A triangle voltage waveform is used in order to drive the laser motor actuator, resulting in a minimal tuning step of 4 pm.

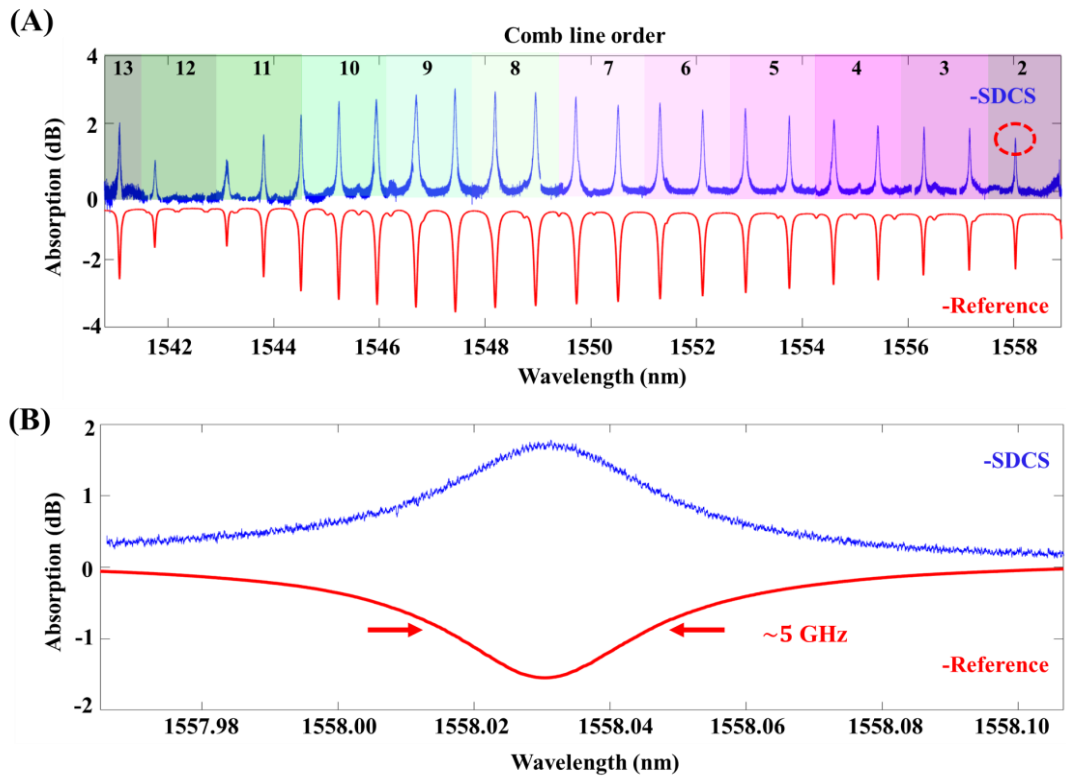

**Figure 3 | Measured molecular absorption spectra.** (A) Absorption spectrum of $2\nu_3$ band of $H^{13}C^{14}N$ measured by SDCS, which is interleaved to cover the entire optical bandwidth; the lower panel shows the reference data, which is NIST-traceable. The pump laser wavelength of SDCS is scanned from 1560.4 nm to 1562 nm. The shaded color regions represent the spectral windows covered by 12 comb lines within the WDM filter bandwidth. (B) The zoomed-in view of the absorption spectrum highlighted in (A): the P20 branch line with SDCS measurements and reference data.

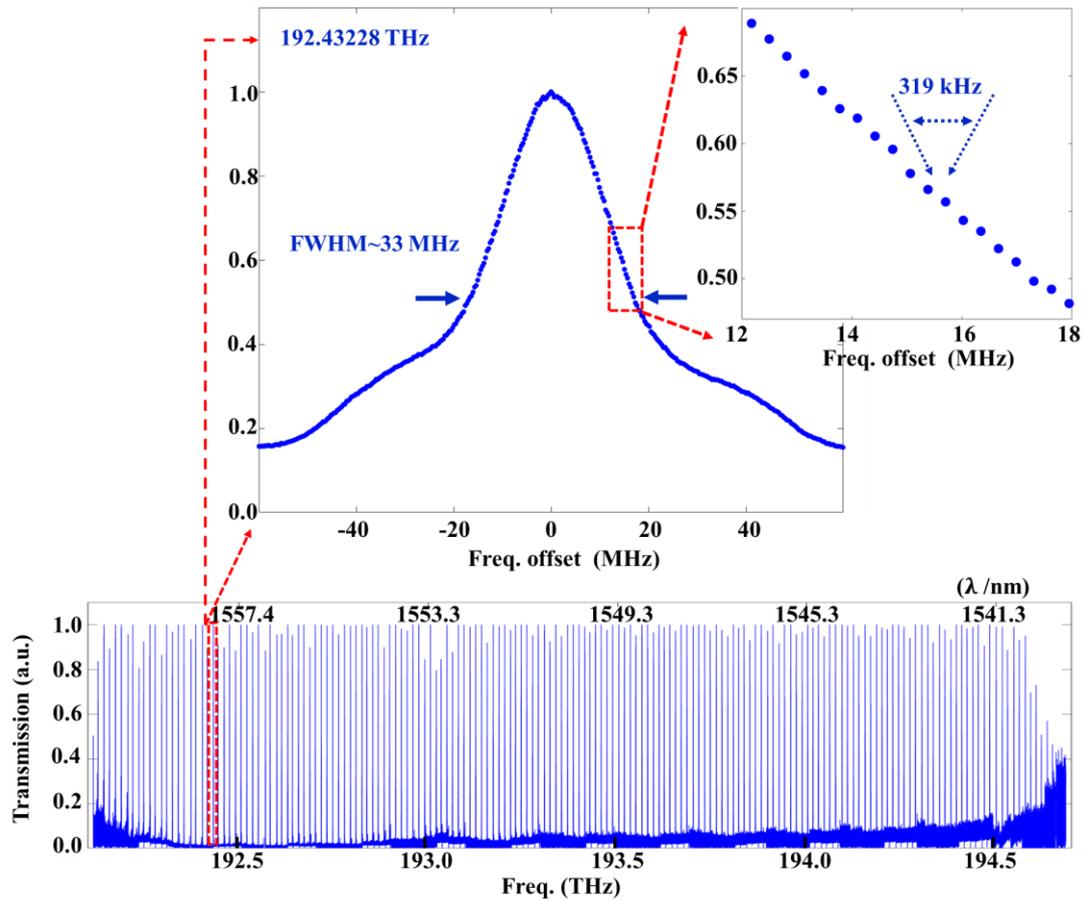

**Figure 4 | Measured Fabry-Perot interferometer transmission spectra.** The bottom panel: The transmission of F-P interferometer using scanning dual-comb spectroscopy technique in the spectral span between 192.1 THz (1560.6 nm) and 194.7 THz (1539.7 nm). The x-axis is mapped from RF domain into optical domain and calibrated by measuring $H^{13}C^{14}N$ simultaneously. The y-axis is normalized in the linear scale. The middle panel: transmission spectrum of the F-P fringe (centered around 1558 nm). The top panel: Higher resolution transmission spectrum around FWHM relative to the peak, each point is sampled every 319±4.6 kHz.